\documentclass[a4paper,12pt]{article}

\usepackage{amsmath}
\usepackage{amssymb}
\usepackage[dvips]{graphicx}
\usepackage[dvips]{psfrag}

\makeatletter
\renewcommand\appendix{\par
  \setcounter{section}{0}%
  \setcounter{subsection}{0}%
  \gdef\thesection{Appendix \@Alph\c@section }
  \renewcommand{\theequation}
  {\Alph{section}.\arabic{equation}}
}
\makeatother

\newcommand{\dd}{{\rm d}}
\newcommand{\ee}{{\rm e}}

\begin{document}

\titlepage

\vspace*{-15mm}   
\baselineskip 10pt   
\begin{flushright}   
\begin{tabular}{r}    
{\tt APCTP-Pre2009-009}\\
July 2009
\end{tabular}   
\end{flushright}   
\baselineskip 24pt   
\vglue 10mm   

\begin{center}
{\Large\bf
 Yet Another Realization\\
of Kerr/CFT Correspondence
}

\vspace{8mm}   

\baselineskip 18pt   

\renewcommand{\thefootnote}{\fnsymbol{footnote}}

Yoshinori~Matsuo\footnote[2]{ymatsuo@apctp.org}, 
Takuya~Tsukioka\footnote[3]{tsukioka@apctp.org} 
and 
Chul-Moon~Yoo\footnote[4]{c\_m\_yoo@apctp.org}

\renewcommand{\thefootnote}{\arabic{footnote}}
 
\vspace{5mm}   

{\it  
 Asia Pacific Center for Theoretical Physics, 
 Pohang, Gyeongbuk 790-784, Korea 
}
  
\vspace{10mm}   

\end{center}

\begin{abstract}
The correspondence between 
the Kerr black hole and a boundary CFT has been conjectured recently.  
The conjecture has been proposed 
first only for the half of the CFT, namely for left movers. 
For right movers, the correspondence has been also found out 
through the suitable asymptotic boundary condition. 
However, the boundary conditions for 
these two studies are exclusive to each other. 
The boundary condition for left movers 
does not allow the symmetry of right movers, and vice versa. 
We propose new boundary condition which 
allows both of left and right movers. 
\end{abstract}

\baselineskip 18pt   

\newpage

Recently, the correspondence between the Kerr black hole 
and a boundary conformal field theory (CFT) was studied~\cite{ghss}.  
They investigated the near horizon geometry which 
had $SL(2, \mathbb R)\times U(1)$ isometries~\cite{baho},  
and considered the asymptotic symmetry 
following the work by Brown and Henneaux~\cite{bh}. 
The Virasoro algebra was realized from an enhancement of the 
rotational $U(1)$ isometry. 
This correspondence yields many generalizations~\cite{many}.  
Soon after an another Virasoro algebra was 
found as an extension of the $SL(2,\mathbb R)$ isometry
\cite{mty}\footnote{
A similar asymptotic symmetry was found in \cite{cd} for warped AdS. 
Constraints in \cite{cd} are weaker than those in \cite{mty}, 
and their symmetries contain the current algebra. 
By using the boundary condition of \cite{mty}, 
we can fix higher order terms of the asymptotic Killing vector. 
This is necessary to see the correspondence of the entropy. 
It should be noticed that the asymptotic charges in \cite{cd} 
are different from those in Kerr black holes, 
because they use $\phi$ as the time coordinate. 
}.
These two symmetries in \cite{ghss} and \cite{mty} correspond to those 
of left and right movers in CFT, respectively. 
In order to access to boundary CFTs, asymptotic boundary conditions 
play a central role. 
However, the asymptotic boundary conditions for 
these two symmetries are not consistent to each other. 
The boundary condition for left movers 
excludes right movers, and vice versa. 
In this letter, we propose new asymptotic boundary condition 
which allows both of left and right movers. 

We start by introducing the Kerr metric in Boyer-Lindquist coordinates: 
\begin{align}
 \dd s^2 &= -\dd t^2 
  + \frac{2mr}{r^2 + a^2 \cos^2\theta} 
  \left(\dd t-a\sin^2\theta \dd\phi\right)^2 
  + \left(r^2 +a^2\right) \sin^2\theta \dd\phi^2 
 \notag\\&\quad
  + \frac{r^2 + a^2 \cos^2\theta}{r^2 -2mr +a^2} \dd r^2 
  + \left(r^2 + a^2\cos^2\theta\right)\dd\theta^2 . 
\end{align}
The parameters $m$ and $a$ are related to the ADM mass 
and the angular momentum as 
\begin{equation}
 M = \frac{m}{G_N} , \qquad\qquad  
 J = \frac{am}{G_N} . 
\end{equation}
The position of the horizon and the Hawking temperature are given by 
\begin{equation}
 r_{\pm} = m \pm \sqrt{m^2 -a^2} , \qquad\qquad 
 T_H = \frac{r_+ - m}{4\pi m r_+} . 
\end{equation}
We consider the near horizon geometry of the Kerr geometry. 
We define new coordinates 
\begin{equation}
 t = 2 \epsilon^{-1} a \hat t , \qquad\qquad
 r = a\left(1 + \epsilon \hat r \right) , \qquad\qquad
 \phi = \hat \phi + \frac{t}{2a},  
\label{Rescale}
\end{equation}
and take the limit of $\epsilon\to 0$ 
to obtain the near horizon geometry. 
For the extremal case $a=m$, the near horizon geometry becomes  
\begin{equation}
 \dd s^2 = - f_0(\theta) \hat r^2 \dd\hat t^2 
  + f_0(\theta) \frac{\dd\hat r^2}{\hat r^2} 
  + f_\phi(\theta)\left(\dd\hat \phi + k \hat r \dd\hat t \right)^2 
  + f_\theta(\theta)\dd\theta^2 , 
\label{NearHorizon}
\end{equation}
with
\begin{equation}
 f_0(\theta) = f_\theta(\theta) = a^2 \left(1+\cos^2\theta\right) , \qquad
 f_\phi(\theta) = \frac{4a^2\sin^2\theta}{1+\cos^2\theta} , \qquad
 k = 1 . 
\label{DetailOfGeometry}
\end{equation}
Hereafter, we consider this near horizon geometry 
and omit `` $\hat\ $ '' of the coordinates. 
The near horizon geometry has $SL(2,\mathbb R)\times U(1)$ isometries  
generated by the following four Killing vectors:  
\begin{subequations}
\begin{align}
 \xi_{-1} &= \partial_t , & 
 \xi_0 &= t \partial_t - r \partial_r , & 
 \xi_{1} &= \left(t^2+\frac{1}{r^2}\right) \partial_t 
 - 2 t r \partial_r - \frac{2k}{r}\partial_\phi , \label{OriginalSL(2,R)}\\
 \xi_\phi &= \partial_\phi , 
\label{OriginalU(1)}
\end{align}
\end{subequations}
where $\xi_{-1}$, $\xi_0$ and $\xi_1$ form the $SL(2,\mathbb R)$, 
and $\xi_\phi$ is the $U(1)$ rotational symmetry. 

Let us consider the asymptotic symmetry of 
this near horizon geometry \eqref{NearHorizon}. 
The asymptotic symmetry is defined by using 
the asymptotic boundary condition. 
For geometries, asymptotic symmetries are 
specified by asymptotic Killing vectors 
which satisfy the Killing equations up to 
the asymptotic boundary condition: 
\begin{equation}
 \pounds_{\xi} g_{\mu\nu} = \mathcal O(\chi_{\mu\nu}),  
  \label{KillingEq}
\end{equation}
where $\pounds_\xi$ is the Lie derivative along $\xi$. 
The metric $g_{\mu\nu} = \bar g_{\mu\nu} + h_{\mu\nu}$ 
contains a small perturbation $h_{\mu\nu}$ from the background
$\bar{g}_{\mu\nu}$  
which is arbitrary but satisfies the asymptotic boundary condition: 
\begin{equation}
 h_{\mu\nu} = \mathcal O(\chi_{\mu\nu}) . 
\end{equation}

Now we propose the following boundary condition: 
\begin{equation}
 h_{\mu\nu} = 
  \bordermatrix{
  & t & r & \phi & \theta \cr
  t 
  & \mathcal O(r^{2}) 
  & \mathcal O(r^{-1})
  & \mathcal O(r^{0}) 
  & \mathcal O(r^{-2}) 
  \cr 
  r 
  & 
  & \mathcal O(r^{-3})
  & \mathcal O(r^{-1})
  & \mathcal O(r^{-3})
  \cr 
  \phi 
  & 
  & 
  & \mathcal O(r^{0})
  & \mathcal O(r^{-1})
  \cr 
  \theta
  & 
  & 
  & 
  & \mathcal O(r^{-1})
  }. 
\label{Constraints}
\end{equation}
The most general form of the asymptotic Killing vector 
which satisfies \eqref{KillingEq} is then given by  
\begin{align}
 \xi &= 
  \Big(
   \epsilon_t(t) + \mathcal O(r^{-1})
  \Big)
  \partial_t 
  + 
  \Big(
   - r \epsilon'_t(t) -r \epsilon'_\phi(\phi) + \mathcal O(r^{0})
  \Big)
  \partial_r 
 \notag\\&\quad
  + 
  \Big(
   \epsilon_\phi(\phi) + \mathcal O(r^{-1})
  \Big)
  \partial_\phi 
  + 
  \Big(\mathcal O(r^{-1})\Big)
  \partial_\theta , 
 \label{AsymptKilling}
\end{align}
where $\epsilon_t(t)$ and $\epsilon_\phi(\phi)$ are arbitrary 
functions of $t$ and $\phi$, respectively. 
By expanding $\epsilon_t(t)$ and $\epsilon_\phi(\phi)$, 
we obtain
\begin{equation}
 l_n = \ee^{-in\phi} \partial_\phi + inr \ee^{-in\phi}\partial_r , 
\qquad\qquad
 \bar l_n = -it^{n+1}\partial_t + i(n+1)r t^n \partial_r . 
\end{equation}
These vectors form the Virasoro algebras:   
\begin{equation}
 i [l_n,l_m] = (n-m) l_{n+m} , \qquad\qquad 
 i [\bar l_n, \bar l_m] = (n-m) \bar l_{n+m} .  
\end{equation}
These two algebras can be understood as 
those of left and right movers in CFT, respectively. 
The rotational Killing vector \eqref{OriginalU(1)} 
is just realized as $l_0$. 
The $SL(2,\mathbb R)$ Killing vectors \eqref{OriginalSL(2,R)} 
can be identified to $\bar l_n$ with $n=-1,\ 0,\ 1$ 
at least for the leading terms. 

We would like to close this letter by making some comments: 

\vspace*{-3mm}

\begin{itemize}
 \item 
 The higher order terms of 
 the asymptotic Killing vector \eqref{AsymptKilling} 
 are not completely arbitrary, i.e.\  
 some of them must not depend on $\theta$. 
 \item 
 We can replace $(r,r)$-component of 
 the asymptotic condition \eqref{Constraints} 
 by $h_{rr} = \mathcal O(r^{-4})$. 
 This asymptotic condition gives almost same vector, 
 but requires the order $1/r$ term in $\phi$-component of 
 the asymptotic Killing vector to vanish. 
 In this case the asymptotic Killing vector is consistent 
 to that in \cite{cd}, but is different from that in \cite{mty}. 
 \item 
 The asymptotic charges and central extensions 
 can be calculated straightforwardly for left movers. 
 For right movers, we have to consider 
 the analytic continuation of $t$. 
 By assuming the correspondence to a holomorphic theory 
 on the complex $t$-plane, 
 the generators can be defined by contour integral on this plane. 
 This definition, however, seems to be different 
 from the interpletation of left and right movers. 
 We need an additional identification between 
 $\phi$ and anti-holomorphic part of $t$, 
 which is still left unclear. 
 \item 
 The analysis of the quasi-local charges done in \cite{mty} 
 is not available for \eqref{AsymptKilling}. 
 For such an analysis, we have to consider higher order corrections 
 of the asymptotic Killing vector, which cannot be fixed by 
 the asymptotic condition \eqref{Constraints}. 
 \item 
 This asymptotic boundary condition cannot be utilized 
 for the warped AdS without $\theta$ direction. 
 One of the constraints comes from 
the $(t,\theta)$-component of \eqref{KillingEq}. 
 This constraint imposes a condition in which 
 $\epsilon_\phi$ does not depend on $t$. 
\end{itemize}

\vspace*{5mm}

\noindent
 {\large{\bf Acknowledgments}}

We would like to thank T.~Nishioka for useful discussions 
and S.~Detournay for stimulating comments. 
This work is supported by YST program in APCTP.

\vspace*{2mm}


\end{document}